\documentclass[fleqn,usenatbib]{mnras}

\usepackage{newtxtext,newtxmath}
\usepackage[T1]{fontenc}

\DeclareRobustCommand{\VAN}[3]{#2}
\let\VANthebibliography\thebibliography
\def\thebibliography{\DeclareRobustCommand{\VAN}[3]{##3}\VANthebibliography}

\usepackage{graphicx}	% Including figure files
\usepackage{subfigure}
\usepackage{float}
\usepackage{url}
\usepackage{amsmath}	% Advanced maths commands
\usepackage{amssymb}	% Extra maths symbols

\title[Photo-z from CSST galaxy flux and image]{Extracting Photometric Redshift from Galaxy Flux and Image Data using Neural Networks in the CSST Survey}

\author[X. Zhou et al.]{
Xingchen Zhou$^{1, 2}$,
Yan Gong$^{1, 3}$$\thanks{E-mail: gongyan@bao.ac.cn}$,
Xian-Min Meng $^{1}$,
Ye Cao $^{1, 2}$,
Xuelei Chen$^{4, 2, 5}$,
Zhu Chen$^{6}$,
Wei Du$^{6}$,\newauthor
Liping Fu$^{6}$,
Zhijian Luo$^{6}$
\\
$^{1}$Key Laboratory of Space Astronomy and Technology, National Astronomical Observatories, Chinese Academy of Sciences, \\20A Datun Road, Beijing 100101, China. \\
$^{2}$University of Chinese Academy of Sciences, Beijing 100049, China \\
$^{3}$Science Center for China Space Station Telescope, National Astronomical Observatories, Chinese Academy of Sciences, \\20A Datun Road, Beijing 100101, China \\
$^{4}$Key Laboratory for Computational Astrophysics, National Astronomical Observatories, Chinese Academy of Sciences, \\20A Datun Road, Beijing 100101, China \\
$^{5}$Centre for High Energy Physics, Peking University, Beijing 100871, China \\
$^{6}$Shanghai Key Lab for Astrophysics, Shanghai Normal University, Shanghai 200234, China\\
}

\date{Accepted XXX. Received YYY; in original form ZZZ}

\pubyear{2015}

\begin{document}

\label{firstpage}
\pagerange{\pageref{firstpage}--\pageref{lastpage}}
\maketitle

\begin{abstract}
    The accuracy of galaxy photometric redshift (photo-$z$) can significantly affect the analysis of weak gravitational lensing measurements, especially for future high-precision surveys. In this work, we try to extract photo-$z$ information from both galaxy flux and image data expected to be obtained by China Space Station Telescope (CSST) using neural networks. We  generate mock galaxy images based on the observational images from the Advanced Camera for Surveys of Hubble Space Telescope (HST-ACS) and COSMOS catalogs, considering the CSST instrumental effects. Galaxy flux data are then measured directly from these images by aperture photometry. The Multi-Layer Perceptron (MLP) and Convolutional Neural Network (CNN) are constructed to predict photo-$z$ from fluxes and images, respectively. We also propose to use an efficient hybrid network, which combines the MLP and CNN, by employing the transfer learning techniques to investigate the improvement of the result with both flux and image data included. We find that the photo-$z$ accuracy and outlier fraction can achieve $\sigma_{\rm NMAD} = 0.023$ and $\eta = 1.43\%$ for the MLP using flux data only, and $\sigma_{\rm NMAD} = 0.025$ and $\eta = 1.21\%$ for the CNN using image data only. The result can be further improved in high efficiency as $\sigma_{\rm NMAD} = 0.020$ and $\eta = 0.90\%$ for the hybrid transfer network. These approaches result in similar galaxy median and mean redshifts ~0.8 and 0.9, respectively, for the redshift range from 0 to 4. This indicates that our networks can effectively and properly extract photo-$z$ information from the CSST galaxy flux and image data.
\end{abstract}


%%%%%%%%%%%%%%%%%%%%%%%%%%%%%%%%%%%%%%%%%%%%%%%%%%

%%%%%%%%%%%%%%%%% BODY OF PAPER %%%%%%%%%%%%%%%%%%

\section{Introduction}\label{sec:introduction}

Galaxy photometric surveys provide an efficient tool to explore the evolution of the Universe and the properties of galaxies. By taking galaxy images in a few photometric bands and deriving the corresponding fluxes, numbers of important information can be analyzed and obtained, such as magnitude, color, morphology, type, size, etc. Then several powerful cosmological probes, e.g. weak gravitational lensing and angular galaxy clustering measurements, can be fulfilled in the cosmological studies. As we know, there are a number of ongoing and next-generation photometric surveys are running and will be started up in near future, e.g. the Sloan Digital Sky Survey (SDSS)\footnote{\url{ http://www.sdss.org/}} \citep{Fukugita96,York00}, Dark Energy Survey (DES)\footnote{\url{https://www.darkenergysurvey.org/}} \citep{Abbott2016,Abbott2021},  the Legacy Survey of Space and Time (LSST) or Vera C. Rubin Observatory\footnote{\url{https://www.lsst.org/}} \citep{Abell2009, Ivezic2019}, the Euclid space telescope\footnote{\url{https://www.euclid-ec.org/}} \citep{Laureijs2011}, and the Wide-Field Infrared Survey Telescope (WFIRST) or Nancy Grace Roman Space Telescope (RST) \footnote{\url{https://www.stsci.edu/roman}} \citep{Green2012, Akeson2019}. These surveys are expected to observe huge amount of galaxies in wide and deep fields, and can provide extremely precise measurements on the cosmic large-scale structure (LSS), properties of dark energy and dark matter, galaxy formation and evolution, and so on.

The photometric redshift (photo-$z$) is a key quantity in photometric surveys. In cosmological studies, accurate photo-$z$ measurements can provide reliable galaxy location information along the line of sight, which is crucial for probing the LSS from the photometric surveys \citep{Zhan2006, Banerji2008}. Besides, the photo-$z$ accuracy is one of the main systematics in weak lensing surveys, and accurate measurements of photo-$z$ and its variance can effectively suppress the systematics and precisely extract the information of kinetic and dynamic evolution of the Universe \citep{Ma2006, Mandelbaum2008, Abdalla2008, Hearin2010}. Currently, there are two main methods used to derive galaxy photo-$z$ from photometric data. One is called template fitting method, which uses typical galaxy spectral energy distributions (SEDs) to fit photometric data and capture spectral features in multi-bands for deriving photo-$z$ \citep{Lanzetta96, Fernandez99, Bolzonella2000}. Another method can be named as training method, and it will extract photo-$z$ by developing empirical relations between redshift and different galaxy properties, such as magnitude, color, and morphology \citep{Collister2004, Sadeh2016, Brescia2021}. This method always can be fulfilled by neural networks using galaxy spectral data with measured spectroscopic redshift as training sample.
Both methods have their own advantages. The template fitting method can be widely used at all redshifts if the selected galaxy SED templates are representative enough for the whole sample. On the other hand, the training method can obtain more accurate photo-$z$ result if the training sample contains sufficiently large and high-quality spectroscopic galaxy data, which can cover the whole redshifts and all features of galaxies in the photometric survey.

Although obtaining qualified training sample would be a challenge for future next-generation photometric surveys that observing deep fields with large redshift coverage, fortunately, a number of high-quality spectroscopic galaxy surveys are currently running or planned, such as Dark Energy Spectroscopic Instrument (DESI)\footnote{\url{https://www.desi.lbl.gov/}} \citep{Levi2019}, Prime Focus Spectrograph \citep[PFS,][]{Tamura16}, Multi-Object Optical and Near-infrared Spectrograph (MOONS)\footnote{\url{https://vltmoons.org/}} \citep{Cirasuolo20,Maiolino20}, 4-metre Multi-Object Spectroscopic Telescope (4MOST)\footnote{\url{https://www.4most.eu/cms/}} \citep{deJong2019}, MegaMapper \citep{Schlegel19}, Fiber-Optic Broadband Optical Spectrograph (FOBOS) \citep{Bundy19}, and SpecTel \citep{Ellis19}. On the other hand, recently the machine learning algorithms, especially neural networks, are remarkably developed. Neural networks are currently widely involved in astronomical and cosmological studies, and numbers of network forms are proposed to solve different problems. For example, there are two widely used networks, i.e. multi-layer perceptron (MLP) and convolutional neural network (CNN). The MLP is the simplest form of neural network, constructed by an input layer, several hidden layers, and an output layer \citep{Haykin1994}, and the CNN proposed by \citet{Fukushima1982} and \citet{Lecun1998}, can extract features from images by learning kernel arrays, and gain great success in detection and recognition tasks.

\begin{figure}
    \centering
    \includegraphics[width=\columnwidth]{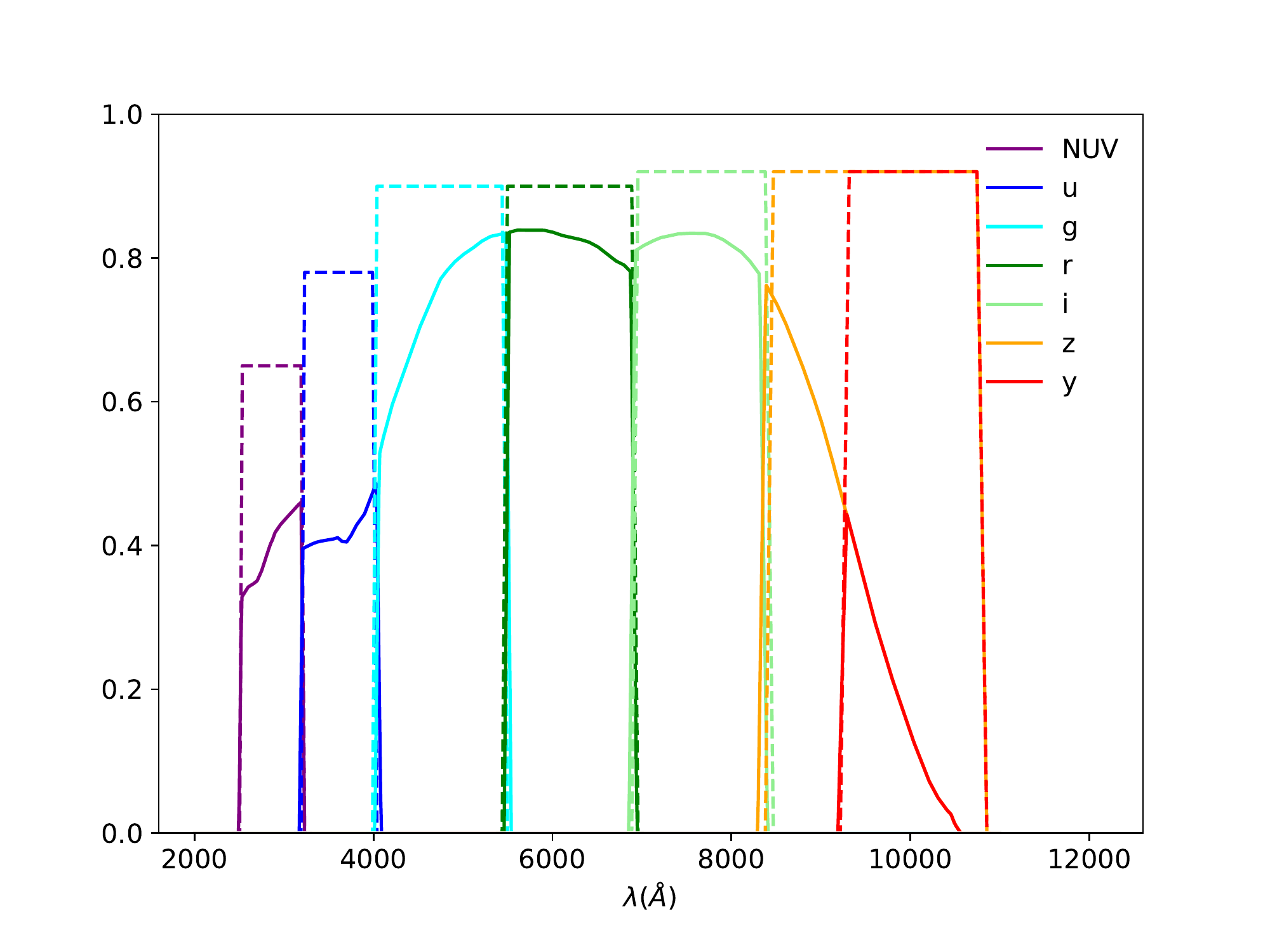}
    \caption{The intrinsic (dashed curves) and real (solid curves) transmissions of the CSST seven photometric filters. The real transmission curves include the effect of detector quantum efficiency. The details of the transmission parameters can be found in \citet{Cao2018} and Meng et al. (in preparation).}
    \label{fig:transmission curve}
\end{figure}

In this work, we focus on the training method by employing neural networks to explore photo-$z$ accuracy in the photometric surveys of the China Space Station Telescope (CSST). The CSST is a next-generation two-meter space telescope, which is planned to launch around 2024~\citep{Zhan2011, Zhan2018, Zhan2021, Cao2018, Gong2019}. It will be in the same orbit with the China Manned Space Station for maintenance and repair. It has seven photometric bands from near-ultraviolet (NUV) to near-infrared (NIR), i.e. $NUV, u, g, r, i, z, y$, covering wavelength range from $\sim250$ nm to $\sim1000$ nm, with 5$\sigma$ magnitude limits as 25.4, 25.4, 26.3, 26.0, 25.9, 25.2 and 24.4, respectively. The intrinsic transmissions and real transmissions including detector quantum efficiency of the CSST seven photometric filters are illustrated in Figure~\ref{fig:transmission curve}. The details of the transmission parameters can be found in \cite{Cao2018} and Meng et al. (in preparation). Totally 17,500 deg$^2$ sky area will be observed during its ten-year mission period, and photometric and spectroscopic surveys will be simultaneously performed. The CSST will answer questions like the properties of dark matter and dark energy, the evolution of the LSS, formation of galaxies, and other important problems in cosmological and astrophysical studies. Many CSST scientific goals and surveys are related to photo-$z$ measurements, e.g. weak gravitational lensing survey, which is the main CSST scientific driver.

In previous studies, the CSST photo-$z$ accuracy has been preliminarily investigated by employing the SED template fitting and MLP neural network methods using the mock flux data directly generated from galaxy SED templates \citep{Cao2018,Zhou2021}. Although relevant background and instrumental noises are properly considered, their results are still ideal, which give photo-$z$ accuracy $\sim0.02-0.03$ and outlier fraction $\sim 3\%$ for the SED fitting method, and $\sim0.01-0.02$ and $\sim0.1\%$ for the neural network. In this work, we will simulate CSST mock galaxy images in the photometric surveys, and measure galaxy flux from these images with aperture photometry technique. Besides, since galaxy morphology should also contain photo-$z$ information ~\citep{Wadadekar2005, Soo2018, WilsonD2020}, we try to estimate photo-$z$ using galaxy images. The MLP and CNN are adopted to extract photo-$z$ from the CSST galaxy flux and image data, respectively. We also propose a hybrid network using the transfer learning technique to efficiently derive photo-$z$ information from both flux and image data by effectively combing the MLP and CNN.

This paper is organized as follows: in Section~\ref{sec:mock data}, we explain the details of generation of the mock image and flux data. In Section~\ref{sec:neural network}, we present architectures of the neural networks we use and details of training process. The results are shown in Section~\ref{sec:result}. We summarize our results in Section~\ref{sec:conclusion}.

\section{Mock Data}\label{sec:mock data}

In order to simulate galaxy images of the CSST photometric survey as real as possible, we generate mock images based on the observations in the COSMOS field performed by the Advanced Camera for Surveys of Hubble Space Telescope (HST-ACS). The image simulation process will be explained in more details by Meng et al. (in preparation). This survey covers about 1.7 deg$^2$ in F814W band, which has similar spatial resolution as the CSST with 80\% energy concentration radius $R_{80}\simeq0.15''$~\citep{Koekemoer2007,Massey2010,Bohlin2016,Cao2018,Gong2019,Zhan2021}. The background noise of the COSMOS HST-ACS F814W survey is also quite low (expected to be $\sim$1/3 of the CSST survey), hence it can provide a good foundation for simulating the CSST galaxy images.

Note that, we only use the measurements from one HST band, i.e. F814W, to generate mock galaxy images without color gradient considered. In real surveys, galaxy morphology will look different in different photometric bands. This would result in wavelength-dependent features and provide more information about galaxy properties, which can be helpful in the neural network training process and improve the accuracy of photo-$z$ estimate. In addition, base on the simulation, the CSST point spread function (PSF) should be more regular or Gaussian-like than the PSF of HST. So the image distortion should be more easily corrected in CSST data processing. However, since the CSST field of view is much larger than the HST (about 300$\times$ larger), precisely understanding and measuring the variances of the PSFs at different locations of the focal plane would be a challenging task, which may affect the image calibration. In this work, for simplicity, we ignore the color gradient and PSF effects on galaxy images.

\begin{figure}
    \centering
    \includegraphics[width=\columnwidth]{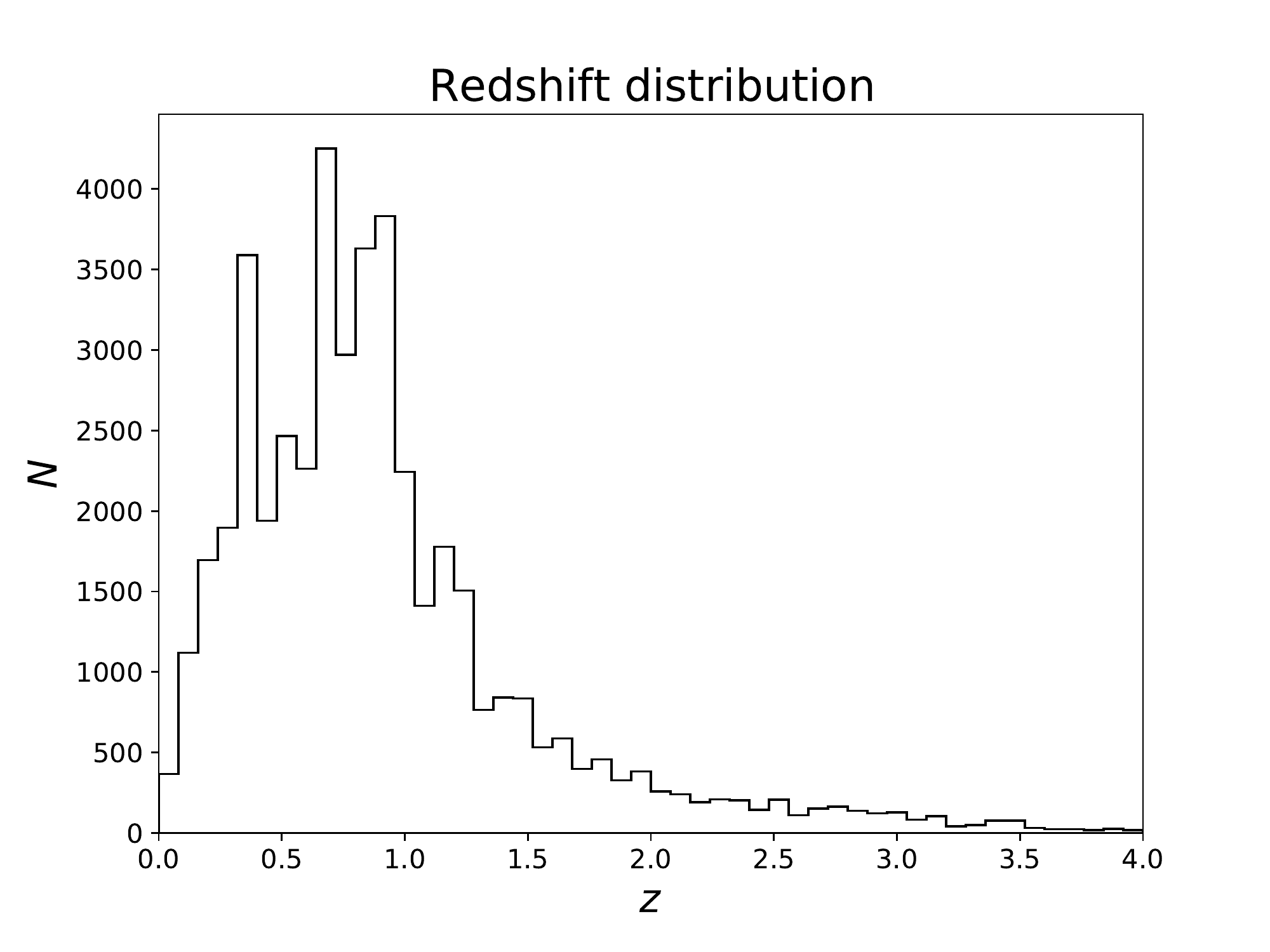}
    \caption{Galaxy redshift distribution of the sample selected from the COSMOS catalog used in the neural networks. These sources have been selected with the SNR greater than 10 in the $g$ or $i$ bands. The distribution has a peak around $z=0.7$, and can extend to $z\sim4$, which is the same as the CSST photometric galaxy sample.}
    \label{fig:redshift distri}
\end{figure}

\begin{figure*}
    \centering
    \includegraphics[width=2.\columnwidth]{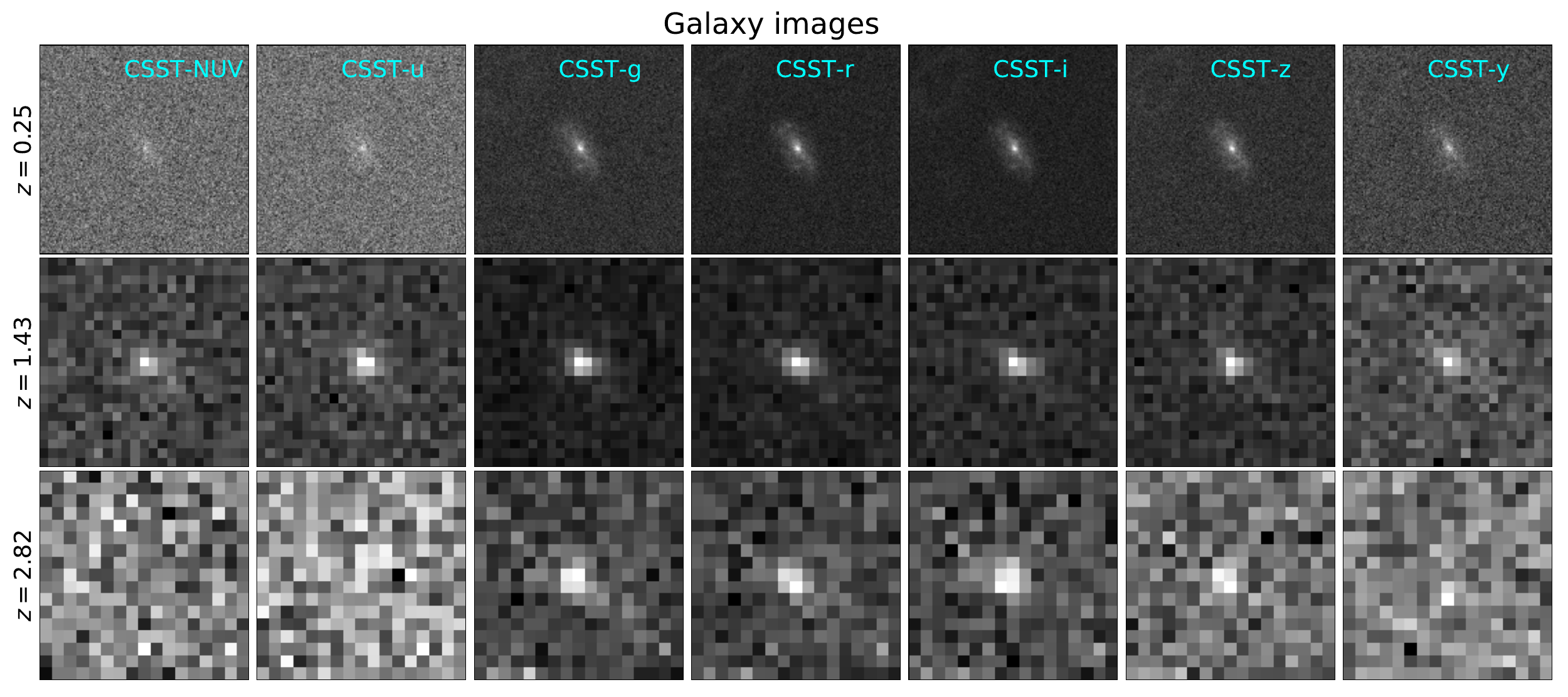}
    \caption{Examples of simulated galaxy images in the seven CSST photometric bands. we notice that noises on the $NUV, u$ and $y$ bands are larger than others, since these bands have lower transmissions. Many sources, especially at high redshifts, are overwhelmed by the background noises in some bands, which may indicate that neural networks are necessarily needed to extract information from these sources.}
    \label{fig:image examples}
\end{figure*}

To obtain high-quality images as "ideal" images for following mock image production, we select the central 0.85$\times$0.85 deg$^2$ area of this survey including $\sim$192,000 galaxies. Then we rescale the pixel size of the COSMOS HST-ACS F814W survey from 0.03$^{\prime\prime}$ to 0.075$^{\prime\prime}$ for matching the CSST pixel size. Next we extract a square stamp image for each galaxy from the survey area, with a galaxy in the center of the stamp image. The size of the square stamp is 15 times of the galaxy's semi-major axis, that means the galaxy stamp images have different sizes. The galaxy's semi-major axis and other relevant galaxy morphological information can be obtained in the COSMOS weak lensing source catalog~\citep{Leauthaud2007}. In addition, we also mask all sources, with signal-to-noise ratio (SNR) >3$\sigma$, in the stamp image except for the galaxy in the center, and replace them by the CSST background noise as we show later, which can be helpful for the neural networks to extract galaxy information.

By making use of galaxy spectral energy distributions (SEDs), we can rescale the galaxy images of the COSMOS HST-ACS F814W survey to the CSST flux level. We first need to find the corresponding SEDs for galaxies by fitting their flux data. Here the COSMOS2015 catalog\footnote{Note that there are still some issues in the COSMOS2015 catalog, and the COSMOS2020 catalog can make relevant improvements and could be used in future analysis~\citep{Weaver2021}.} is adopted to match galaxies in HST-ACS F814W survey, which contains 220,000 galaxies with measures of galaxy redshift, magnitude, size, and so on~\citep{Laigle2016}. To avoid additional errors, we use the 31 SED templates from $\it LePhare$ code~\citep{Arnouts1999, Ilbert2006}, which are also used in the COSMOS2015 catalog, to fit the galaxy flux data. Since the CSST has large wavelength coverage from NUV to NIR, for galaxies detected at $z\gtrsim2$, we also extend the wavelength range of these templates from $\sim900$ \r{A} to $\sim90$ \r{A} using the BC03 method \citep{Bruzual03}. The details can be found in \cite{Cao2018}. We select high-quality  photometric data with reliable photo-$z$ measurement and about 100,000 galaxies are selected. Besides, when fitting the flux data with the SED templates and finding the best-fit SEDs, we also include dust extinction and emission lines, such as Ly$\alpha$, H$\alpha$, H$\beta$, [OII] and [OIII].

After obtaining the best-fit SED for each galaxy, we can calculate the theoretical flux data observed by the CSST. The theoretical flux in $\rm e^- s^{-1}$ or electron counting rate for a band $j$ can be estimated as
\begin{equation}
    C^j_{\rm th} = A^j_{\rm eff} \int S(\lambda) \tau_j(\lambda) \frac{\lambda}{hc} d\lambda.
\end{equation}
Here $A^j_{\rm eff}$ is the effective telescope aperture area, $h$ and $c$ are the Planck constant and speed of light, $S(\lambda)$ is the galaxy SED, and $\tau_j(\lambda)=T_j(\lambda)Q_j(\lambda)M_j(\lambda)$ is total system throughput, where $T_j(\lambda)$, $Q_j(\lambda)$, and $M_j(\lambda)$ are the intrinsic filter transmission, detector quantum efficiency, and total mirror efficiency, respectively \citep{Cao2018,Zhou2021}. Then the theoretical flux electron counts in $j$ band can be expressed as
\begin{equation}
    F^j_{\rm th} = C^j_{\rm th}\, t_{\rm exp} N^j_{\rm exp}\, ,
\end{equation}
where $t_{\rm exp}=150$ s is the exposure time, and $N^j_{\rm exp}$ is the number of exposure that $N^j_{\rm exp}=2$ for $u$, $g$, $r$, $i$ and $z$ bands, and 4 for $NUV$ and $y$ bands.
For the measured galaxy flux $F^j_{\rm obs}$ in the HST-ACS F814W survey, we collect the flux of galaxy in the center of stamp image with photometry aperture size of 2 times of the Kron radius \citep{Kron1980}, i.e. $2\times R_{\rm Kron}$. Then we can rescale the measured galaxy flux to the expected flux observed by the CSST, and produce the CSST mock galaxy images in a band. Note that, in the meantime, we also rescale the background of the stamp image by the same factor, i.e. rescaling the whole stamp image including both galaxy in the center and background, which can be used in the following background correction.

Besides rescaling galaxy flux, we also adjust the background noise in the stamp image to the CSST level. Since the background noise in the HST-ACS F814W survey is expected to be much lower than the CSST survey, we need to add additional noise into the image in $j$ band, and we have
\begin{equation}
    N_{\rm add}^j = \sqrt{\left(N_{\rm bkg,th}^j\right)^2 - \left(N_{\rm img}^j\right)^2},
    \label{eq:N_add}
\end{equation}
where $N_{\rm img}^j$ is the background noise of the rescaled image discussed above, and $N_{\rm bkg,th}^j$ is the theoretical CSST background noise per pixel, which can be estimated by
\begin{equation}
    N_{\rm bkg,th}^j =  \sqrt{(B^j_{\rm sky} + B_{\rm dark})t_{\rm exp}N^j_{\rm exp} + R_{\rm n}^2 N^j_{\rm exp}}.
    \label{eq:N_bkg}
\end{equation}
Here $B_{\rm dark}=0.02\ \rm e^{-}s^{-1}pix^{-1}$ is the dark current, $R_{\rm n}=5\ \rm e^{-}pix^{-1}$ is the read-out noise, and $B^j_{\rm sky}$ is the sky background in unit of $\rm e^{-}s^{-1}pix^{-1}$, which is given by
\begin{equation}
    B^j_{\rm sky} = A^j_{\rm eff} l_{\rm pix}^2 \int I_{\rm sky}(\lambda) \tau_j(\lambda) \frac{\lambda}{hc} d\lambda,
    \label{eq:B_sky}
\end{equation}
where $l_{\rm pix}$ is the pixel size in arc-seconds, and $I_{\rm sky}(\lambda)$ is the surface brightness intensity of the sky background in units of $\rm erg\,s^{-1}cm^{-2}${\AA}$^{-1}{\rm arcsec}^{-2}$. We evaluate $B^j_{\rm sky}$ based on the measurements of the earthshine and zodiacal light for the `average' sky background case given in~\cite{Ubeda2011}. We find that $B^j_{\rm sky}$ are 0.0023, 0.018, 0.142, 0.187, 0.187, 0.118 and 0.035 e$^-$s$^{-1}$pix$^{-1}$ for $NUV$, $u$, $g$, $r$, $i$, $z$ and $y$ bands, respectively, which are consistent with the results in \cite{Cao2018}. Then $N_{\rm bkg,th}^j$ are calculated as 10.65, 7.84, 9.93, 10.59, 10.59, 9.56 and 11.53 $\rm e^{-}$ for these bands. Thus the additional noise $N_{\rm add}^j$ can be calculate by subtracting the rescaled image noise $N_{\rm img}^j$ as shown in Equation \ref{eq:N_add}, and added to pixels in stamp images by sampling from a Gaussian distribution with mean$=0$ and $\sigma=N_{\rm add}^j$ in band $j$. Then we obtain the final mock CSST galaxy images for each band.

Next we measure galaxy flux data from the CSST mock galaxy images using aperture photometry method. Firstly, to obtain high SNR source detections and morphological measurements, we stack the $g$, $r$, $i$, and $z$ band images for each galaxy to create the detection images. Secondly, we measure the Kron radius along galaxy major- and minor-axis  to find an elliptical aperture with the size $1\times R_{\rm Kron}$, that can improve the SNR. Finally, the flux and error in $j$ band can be calculated from electron number $N^{j}_{\rm e^-}$ and error $\sigma^{j}_{\rm e^-}$ measured within the aperture. Note that if a galaxy in a band is very faint, the measured flux could be negative due to background noise. This effect will not affect the training process of our neural networks, and as we show in Section~\ref{sec:neural network}, we will rescale it and reserve the information. The SNR of measured galaxies in band $j$ can be estimated as
\begin{equation}
    {\rm SNR}_j = \frac{F^j_{\rm obs}}{\sqrt{F^j_{\rm obs}+N_{\rm pix}(B_{\rm sky} + B_{\rm dark})t_{\rm exp}N^j_{\rm exp} + R_{\rm n}^2 N^j_{\rm exp}N_{\rm pix}}},
\end{equation}
where $F^j_{\rm obs}$ is the observed electron counts, and $N_{\rm pix}$ is the number of pixels covered by a galaxy for the CSST, which can be derived from the COSMOS2015 catalog.

\begin{figure}
    \centering
    \includegraphics[width=\columnwidth]{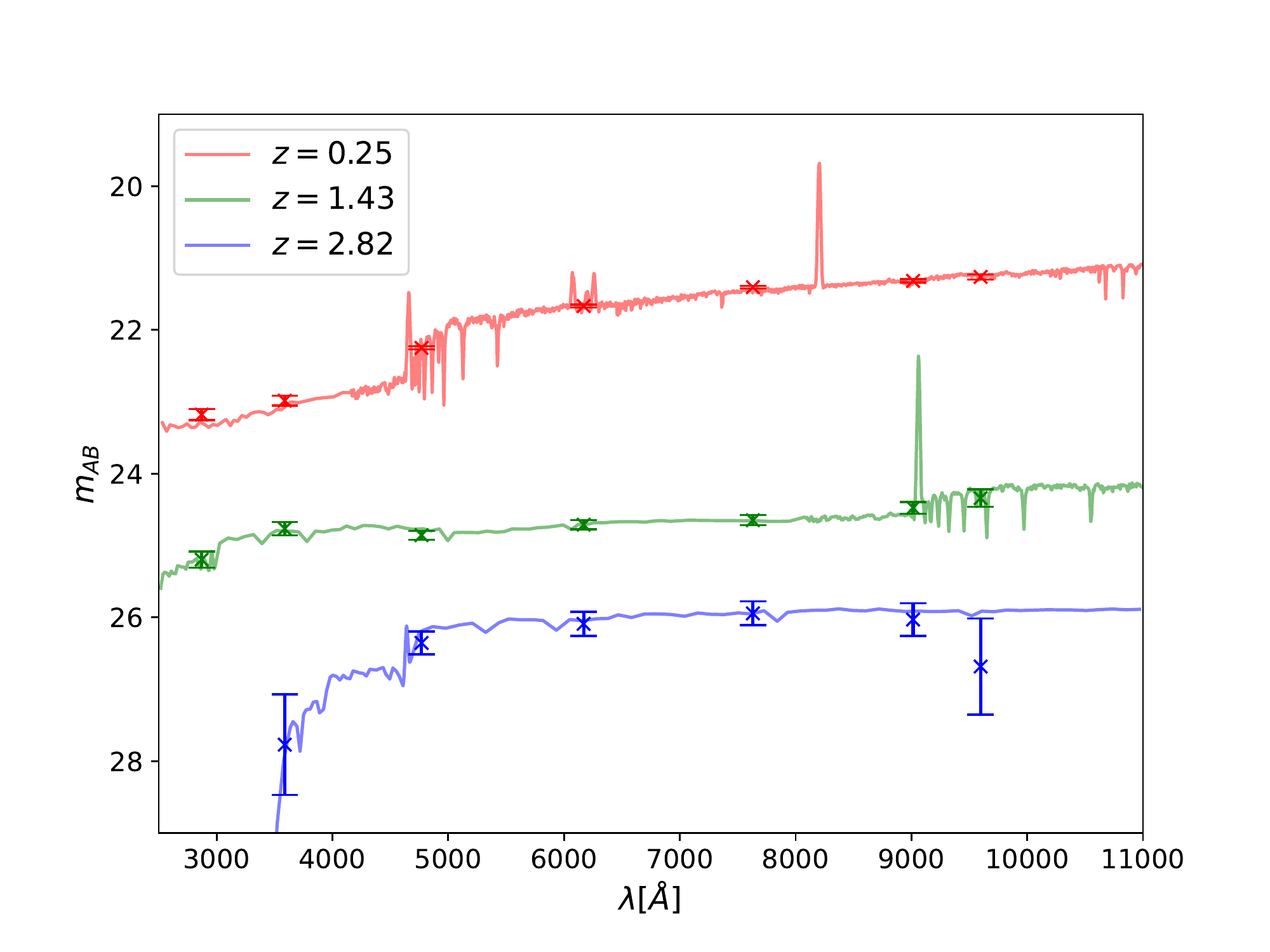}
    \caption{The flux data directly measured from the mock CSST galaxy images shown in Figure~\ref{fig:image examples} by aperture photometry in the seven photometric bands. The SEDs of the corresponding galaxy images are also shown in solid curves as comparison.}
    \label{fig:sed examples}
\end{figure}

 In Figure~\ref{fig:redshift distri}, we show the redshift distribution of the galaxy sample selected from the COSMOS catalog (the selection details can be found in the next section).  Since the selected galaxy sample has high-quality photo-$z$ measurement, we assume that these photo-$z$s can be seen as spectroscopic-redshifts (spec-$z$s), that can be used in the neural networks for method validation purpose. We can see that the distribution has a peak around $z=0.6-0.7$, and can extend to $z\sim4$, which is consistent with the CSST galaxy redshift distribution in previous studies \citep{Cao2018,Gong2019,Zhou2021}. In Figure~\ref{fig:image examples}, the CSST mock galaxy stamp images at different redshifts have been shown. We can find that the low-$z$ galaxies have higher SNR with low backgrounds, while high-$z$ galaxies can be dominated by the background noise, especially in the bands with low transmissions. This indicates that it may be necessary to employ the machine learning technique to extract information from these bands. The corresponding measured galaxy fluxes from mock images by aperture photometry in the seven CSST bands are shown in Figure~\ref{fig:sed examples}. The SEDs are rescaled to the levels of flux data as comparison. As we show, our measured mock flux data match the corresponding SEDs very well, and can correctly represent the features.

\section{Neural Networks}\label{sec:neural network}

The MLP and CNN are used to derive the photo-$z$ from the mock flux and image data obtained in the last section, respectively. We also explore the effective way of combining these two networks for testing the improvement of the photo-$z$ accuracy by including both flux and image data.  The details of architectures and training processes of these networks are discussed in this section. All networks we construct are implemented by Keras\footnote{\url{https://keras.io}} with TensorFlow\footnote{\url{https://www.tensorflow.org}} as backend.

\subsection{Network Architecture}\label{sec:net arch}
The flux data of a galaxy observed by the CSST would contain seven discrete data points in the seven CSST photometric bands, and we adopt the MLP to predict photo-$z$ from these data.
The MLP is the simplest form of neural networks, consisting of an input layer, several hidden layers and an output layer. Every layer has several units or perceptrons, and the hidden layers can learn the internal relationship between input data points. Layers are connected by weights and biases, which can be initialized by random distributions and optimized in the training process.

We use the measured band flux, color, and error as the inputs of the MLP \citep{Zhou2021}. In order to speed up the training process, we need to rescale these data and make the network effectively extract the relevant information. A useful way we find for rescaling is to take the logarithm of these data. However, as we discussed in the last section, since the flux and error are measured by the aperture photometry and because of the fluctuations of noises, we may encounter some negative fluxes, especially in the CSST $NUV$, $u$, and $y$ bands with relatively low transmissions. To fix this issue, we propose to use the following function
\begin{equation}
    f(x) = \left\{
    \begin{array}{lcc}
        \log(x)   &  & {x > 0,} \\
        -\log(-x) &  & {x < 0.} \\
    \end{array} \right.
    \label{eq:scale function}
\end{equation}
This function can naturally convert a negative value to be a logarithmic one, and could effectively reserve its information which can be useful to derive photo-$z$ as we find \footnote{We also try to set the negative flux to be zero or use its measured upper-limit information, but we find that the negative flux also contains useful information when deriving photo-$z$, and we should reserve it in the rescaling process.}. Before rescaling, we perform normalization-like process for the fluxes and errors to speed up the training process of the network. For the flux data, we divide it by a fixed value that is close to the measured flux. Here we take this fixed value as the flux-limit for each CSST band, which is derived from the magnitude limit shown in Section~\ref{sec:introduction}. Note that our result is not sensitive to the selection of this fixed value in each band. For the error, we divide it by its corresponding flux to obtain a relative error. To include more information, we also construct a color-like quantity, which is obtained by calculating the ratio of fluxes in the two bands. Relative error and color-like quantity are also rescaled by Equation~\ref{eq:scale function}. For simplicity, we call these rescaled normalized flux, relative error, and color-like quantity as flux, error, and color in our following discussion.

\begin{figure*}
    \centering
    \includegraphics[scale=0.37]{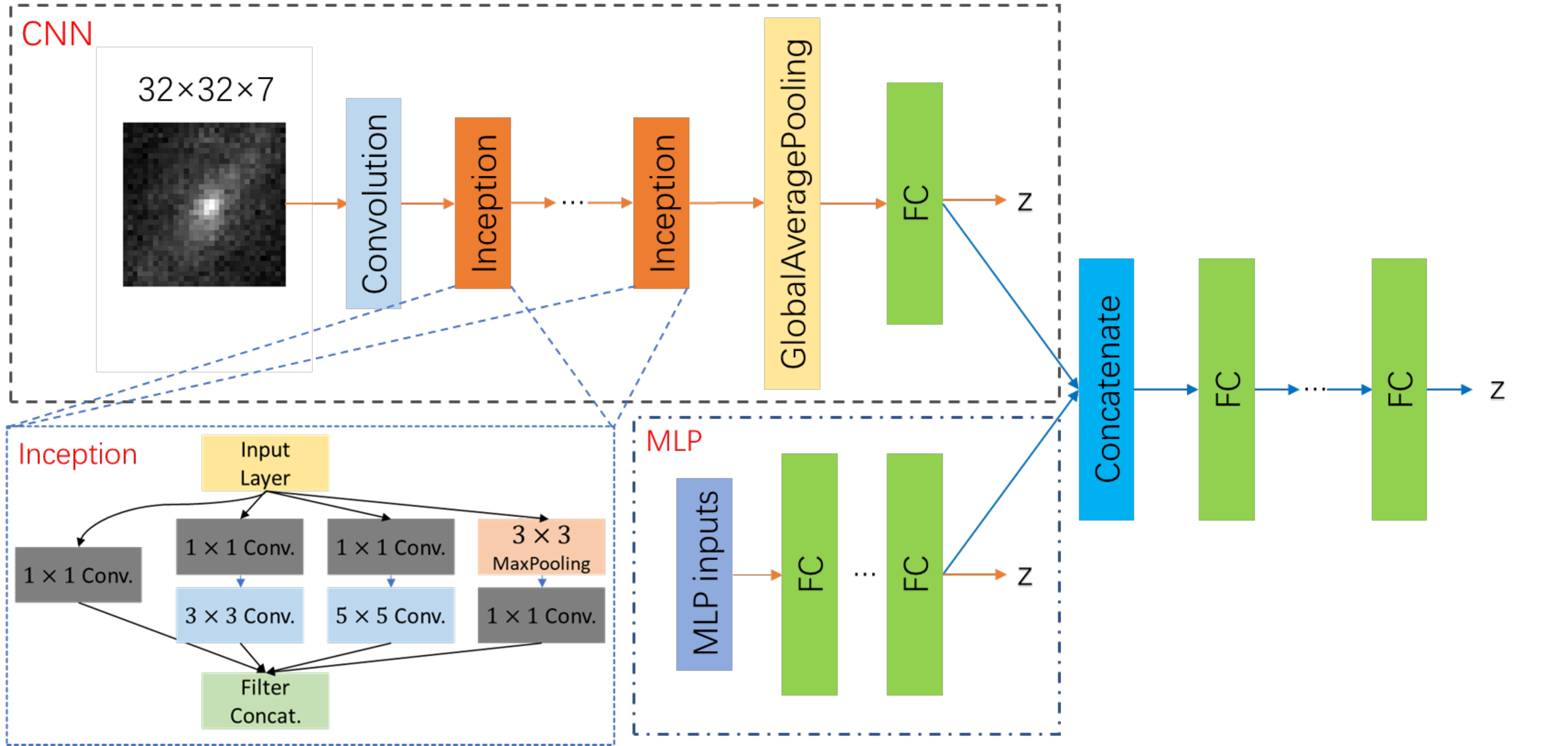}
    \caption{Architecture of the MLP, CNN and Hybrid networks. The MLP structure is shown in blue dash-dotted box. The inputs are rescaled fluxes, colors and errors, and totally 6 hidden layers are structured. The CNN structure is displayed in the dashed black box. The inputs are $(32\times32\times7)$ images, convolved and downsampled by the convolutional layer. Then three inception blocks are structured to obtain the features of size 2, which is flattened by global average pooling to a vector of size 72. Next the fully-connected layer with 40 units is applied, and then the photo-$z$ can be obtained. The inception blocks are illustrated in the dashed blue box. We use $(3\times3)$ and $(5\times5)$ kernels to extract features parallelly, and $(1\times1)$  kernels are adopted to reduce the number of features for increasing efficiency of computation. The hybrid network are constructed by concatenating the vectors extracted by the MLP and CNN from its fully-connected layer. Totally 6 fully-connected layers with 80 units are structured, and after each layer the BatchNormalization and ReLU activation function are applied.}
    \label{fig:network architecture}
\end{figure*}

Therefore, our MLP has 20 inputs, which contain seven fluxes, seven errors, and six colors. We construct 6 hidden layers with 40 units or neurons in each layer, that means a classic structure with $n:2n:...\ 2n:1$ is applied where $n$ is number of data elements. We find that less number of hidden layers cannot get accurate results, while more layers result in lower efficiency and would not further improve the accuracy. Except for the first hidden layer, the BatchNormalization layer is applied to reduce overfitting problem~\citep{Ioffe2015}, and every hidden layer is activated by the Rectified Linear Unit (ReLU) nonlinear function with a form $y={\rm max}(0,x)$~\citep{Nair2010}. The outputs of the MLP is the photometric redshift. The details of the architecture of the MLP are shown in the blue dash-dotted box of Figure~\ref{fig:network architecture} and Table~\ref{tab:MLP_param}.

We adopt the CNN to analyze galaxy images. The mock CSST galaxy images observed in the seven photometric bands can be seen as 2d arrays or matrices in seven channels. The CNN can process multiple-channel arrays, which is suitable in our case for extracting photo-$z$ from galaxy images. Note that the galaxy images we obtain are in different sizes due to the stamp slicing procedure based on semi-major axis mentioned in Section~\ref{sec:mock data}, but the CNN can only process images in the same sizes. So we crop centrally the images larger than a size of threshold $S_{\rm threshold}$ and pad the images smaller than $S_{\rm threshold}$ with random normal noise. Padding with random noise instead of 0 or fixed values can mimic the observations more realistically. This random normal noises are typical backgrounds derived from Equation~\ref{eq:N_bkg} in Section~\ref{sec:mock data}. The $S_{\rm threshold}$ is set to be 32 pixels in our work, and other sizes with $S_{\rm threshold}=16$ and 64 are also explored. We find that smaller $S_{\rm threshold}$ loses too much information since most galaxies occupy pixels larger than 16, and larger $S_{\rm threshold}$ will be dominated by random padding noises that makes the CNN cannot effectively concentrate on extracting information from galaxy in the center of image. We also make use of the inception blocks to extract image features parallelly~\citep{Szegedy2014}. The inception block is widely used for predicting photometric redshift through galaxy images, and the networks constructed with it display excellent performance~\citep[see e.g.][]{Pasquet2019,Henghes2021}. Inception block is illustrated in dashed blue box in Figure~\ref{fig:network architecture}. This block uses kernels of different sizes, i.e. $(3\times3)$ and $(5\times5)$, to extract image features parallelly. The $(1\times1)$ kernels are used to reduce number of channels to increase computation efficiency.

%\begin{figure}
%	\includegraphics[width=\columnwidth]{MLP_archi.jpeg}
%    \caption{The architecture of the MLP used to analyze the mock CSST flux data. Inputs are rescaled fluxes, colors and errors. Totally 6 hidden layers are structured, and the output is the value of photo-$z$.}
%    \label{fig:mlp archi}
%\end{figure}

\renewcommand{\arraystretch}{1.5}
\begin{table}
    \caption{Details of MLP architecture.}
    \label{tab:MLP_param}
    \begin{center}
        \begin{tabular}{lcc}
            \hline
            Layers             & Output Status$^a$ & Number of params.$^b$ \\
            \hline
            \hline
            Input              & 20                & 0                     \\
            \hline
            FC$^c$             & 40                & 840                   \\
            \hline
            ReLU               & 40                & 0                     \\
            \hline
            FC$^c$             & 40                & 1640                  \\
            \hline
            BatchNormalization & 40                & 160$^d$               \\
            \hline
            ReLU               & 40                & 0                     \\
            \hline
                               & ...$^e$           &                       \\
            \hline
            Output             & 1                 & 41                    \\
            \hline
            \hline
        \end{tabular}
    \end{center}
    \vspace{-2mm}
    \textbf{Notes}.\\
    $^a$ Number of data points or neurons. \\
    $^b$ Total number of parameters: 9,881. \\
    $^c$ FC: fully connected layer. \\
    $^d$ Half of them are non-trainable parameters.\\
    $^e$ 4 repeats of FC + BatchNormalization + ReLU. \\
\end{table}

\renewcommand{\arraystretch}{1.5}
\begin{table}
    \caption{Details of CNN architecture.}
    \label{tab:CNN_param}
    %\centering
    \begin{center}
        \begin{tabular}{lcc}
            \hline
            Layers               & Output Status$^a$ & Number of params.$^b$ \\
            \hline
            \hline
            Input                & (32, 32, 7)       & 0                     \\
            \hline
            Conv2D               & (16, 16, 32)      & 2048                  \\
            \hline
            BatchNormalization   & (16, 16, 32)      & 128$^c$               \\
            \hline
            ReLU                 & (16, 16, 32)      & 0                     \\
            \hline
            Inception            & (8, 8, 72)        & 9824                  \\
            \hline
            Inception            & (4, 4, 72)        & 11744                 \\
            \hline
            Inception            & (2, 2, 72)        & 11744                 \\
            \hline
            GlobalAveragePooling & 72                & 0                     \\
            \hline
            FC$^d$               & 40                & 2920                  \\
            \hline
            ReLU                 & 40                & 0                     \\
            \hline
            Output               & 1                 & 41                    \\
            \hline
            \hline
        \end{tabular}
    \end{center}
    \vspace{-2mm}
    \textbf{Notes}.\\
    $^a$ Format: (dimension, dimension, channel) or number of neurons.\\
    $^b$ Total number of parameters: 38,449.\\
    $^c$ Half of them are non-trainable parameters.\\
    $^d$ FC: fully connected layer.\\
\end{table}

In our CNN, the inputs are $(32\times32\times7)$ images. Firstly, we apply a convolutional layer with 32 kernels of size $(3\times3)$ and stride size 2 to extract features and downsample images to 16. Then we structure 3 inception blocks obtaining feature images of size 2. After these blocks, we use global average pooling to vectorize these features to 72 values~\citep{Lin2013}, and then apply a fully-connected layer of 40 units. Finally, the network outputs are photo-$z$ values. In addition, after each convolutional layer, BatchNormalization layer and LeakyReLU activation function are applied~\citep{Maas2013}. The architecture of our CNN are illustrated in dashed black box in Figure~\ref{fig:network architecture}, and the details of the CNN architecture are shown in Table~\ref{tab:CNN_param}.

Since the CNN can extract photo-$z$ from the galaxy image mainly using morphology information, it is necessary to explore the improvement of the photo-$z$ accuracy using both galaxy flux and image data.  Hence, we construct a hybrid network to combine the MLP and CNN for including all data, i.e. galaxy fluxes, errors, colors, and images. In the hybrid network, the CNN and MLP parts are the same as the ones described above. Their fully connected layers are concatenated, and this means the features extracted by the CNN and MLP from galaxy images and flux are combined together. The constructed vector by the combination is of size 80 (i.e. 40 from the CNN and 40 from the MLP). Then we structure 6 fully connected layers with 80 units in each layer, and the BatchNormalization and ReLU activation function are applied after each layer. Then the network outputs the photo-$z$ values. The illustration of the hybrid network is shown in Figure~\ref{fig:network architecture}.

\subsection{Training}\label{sec:training}

The photo-$z$ data of the CSST will be mainly used in the analysis of the weak gravitational lensing surveys, which needs high-quality sample with accurate photo-$z$ information. Therefore, following previous works \citep{Cao2018,Zhou2021}, we select approximately 40,000 high-quality sources with the SNR in $g$ or $i$ band larger than 10 from the original data set (containing $\sim$100,000 galaxies). So these 40,000 sources have both image and flux data, and also are assumed to have accurate spec-$z$ measurements, although currently they only have high-quality photo-$z$ measurements in the COSMOS catalog. In the future real CSST survey, we will use the real spec-$z$ sample obtained from future spectroscopic surveys as the training sample. We should notice that this selection criterion is conservative, and large amounts of galaxies with low SNRs will be removed, which can be used in weak lensing survey. However, since the systematical errors are the main problem to affect the accuracy of next-generation weak lensing measurements, and photo-$z$ uncertainty is one of the main components, we will try to suppress the systematics with high priority, even losing a fraction of galaxies. Since the quality of galaxy shape measurement also should be considered as an important criterion in weak lensing surveys, other potentially better selection criterions will be explored in future works. Then we divide the above sample into the training and testing data, and randomly save 10,000 sources for testing and the rest of 30,000 data for training, which means the ratio of the training and testing data is roughly $3: 1$. In order to investigate the effect of the number of training data on the photo-$z$ estimation, we also try ratios roughly $1: 1$ and $1:3$ for the training and testing samples.

Note that we have assumed all the samples we use to train the networks have measured spec-$z$ provided by future spectroscopic surveys as discussed in the introduction. Galaxies in these samples can fully represent the features of galaxies in the CSST photometric survey, such as the redshift and magnitude distributions, galaxy types, and so on. In real surveys, this assumption may not hold, and selection effect of spectroscopic samples will influence the estimates on photo-$z$. Since the the result derived from the neural networks is highly dependent on the training sample, if there is a kind of galaxies that is not covered by the spec-$z$ training set, the networks probably would not provide proper photo-$z$ estimates for these galaxies. However, we may fix this problem by analyzing the photometric and spectroscopic samples and finding the missing galaxies in the spec-$z$ training set by some means, e.g. adopting the self-organizing map suggested in~\citet{Masters2015}. Then we can try to cover these galaxies on purpose with the help of available spectroscopic surveys.

\begin{figure}
    \centering
    \includegraphics[width=\columnwidth]{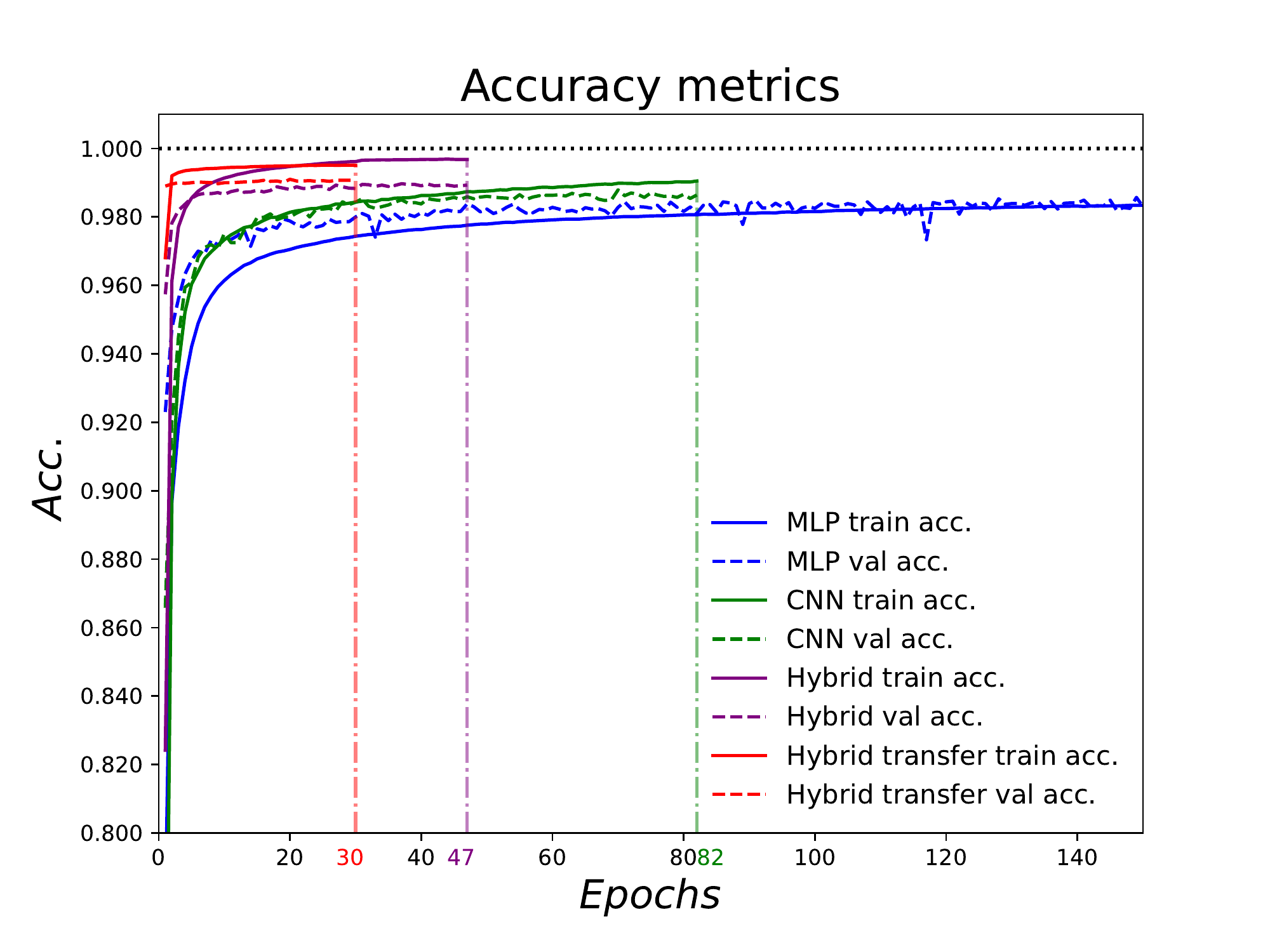}
    \caption{Accuracy metrics of different networks in training (solid) and validation (dashed). Blue, green, purple, and red lines represent MLP, CNN, Hybrid, and Hybrid transfer respectively. Validation accuracy reaches approximately 0.98 for the MLP. For CNN, the training stops at the 82nd epoch, since the validation accuracy does not improve in previous 10 epochs, and the maximum accuracy can be higher than MLP ones. Hybrid and Hybrid transfer networks can reach highest accuracy trained only 47 and 30 epochs, respectively. The vertical dotted lines shows the epochs when training stops.}
    \label{fig:acc}
\end{figure}

In our MLP, we use the mean absolute error (MAE) as our loss function, which calculates the average of absolute difference between predicted redshifts and true redshifts. The Adam optimizer is adopted to optimize learnable weights~\citep{Kingma2014}. Adam optimizer is an efficient stochastic optimization method requiring only first-order gradients, and it can adjust learning rate automatically in the training process for every trainable weight in network. In addition, we define an accuracy metric based on outlier percentage defined as $|z_{\rm true} - z_{\rm pred}| < 0.15(1 + z_{\rm true})$, and the network will monitor this metric when training. We set the initial learning rate to be $10^{-4}$, and use ModelCheckpoint callback to save the model with highest validation accuracy. The maximum number of epochs is set to be 150 in the training process. In order to reduce the effect of the statistical noise, the MLP training data are augmented by random realizations based on flux errors with the Gaussian distribution. This method has found to be effective for eliminating the effect of statistical noise in the training process~\citep{Zhou2021}. Here we use 50 realizations for every training source. We find that more realizations do not improve the results significantly. In Figure~\ref{fig:acc} we show the accuracy versus epoch for both training and validation samples in blue solid and dashed curves, respectively. We can see that the validation accuracy can reach $\sim$0.98. The validation samples are mainly used to tune the hyper-paramaters of a network. On the phase of tuning networks, we randomly select validation sample as 10\% of training sample.

In the CNN, we also use MAE and Adam optimizer with initial learning rate setting to be $10^{-4}$, and the learning rates are reduced by 10 through ReduceLROnPlateau callback when validation accuracy does not improve in 5 epochs. To make the training efficient, we use EarlyStopping callback to shut down training if no improvement occurs in 10 epochs, and save the model with the highest validation accuracy. To augment the training data, galaxy images for training are rotated and flipped, resulting in $8\times$ original data size. In Figure~\ref{fig:acc}, the green lines show the accuracy curves for the CNN training (solid) and validation (dashed) samples. We can find that they stops at 82nd epoch, meaning previous 10 epochs have no improvement in validation accuracy, and the model is saved at the 72nd epoch.

For the hybrid network, settings of the loss function, optimizer and callbacks are the same as the corresponding CNN described above. The hybrid network inputs include both flux and image data for a galaxy, i.e. flux, error, and color for the MLP and a corresponding galaxy image for the CNN. The inputs of the MLP part in the hybrid network are also augmented by 50 realizations based on flux errors. For the CNN part, we need to input the corresponding galaxy image given a random flux realization. To make the training process efficient and extracting more information, we input a galaxy image by randomly rotating or flipping. This means that, for a galaxy, the inputs of the hybrid network contain 50 realizations of the flux data and the corresponding 50 randomly rotating or flipping galaxy images.

To investigate the possibility that if we can further improve the network efficiency and accuracy, we also propose and explore a hybrid transfer network, which is inspired by the techniques from the transfer learning. Transfer learning is a technique focusing on using knowledge gained from one problem to solve a different but related problem~\citep{West2007}. Here we borrow this idea and combine the features learned by MLP and CNN together. As we show in the following discussion, this network can have better training efficiency and obtain higher photo-$z$ accuracy. Since our MLP and CNN have already been well trained and can provide good photo-$z$ accuracies, we can directly freeze the weights obtained by the MLP and CNN and transfer them to the MLP and CNN parts of the hybrid network, i.e. constructing a hybrid transfer network. In this way, the MLP and CNN parts are restricted to learn features that is useful to predict photo-z, and the concatenated features in the joint fully connected layers can improve the estimation. Note that the  architecture of the hybrid transfer network is identical to the hybrid network, except for applying a different training strategy inspired by transfer learning. Besides, in the MLP and CNN parts, we just freeze the first five fully connected layers in the MLP and layers before the global average pooling layer in the CNN, and the last fully connected layers in the MLP part and the one in the CNN part are still involved in the training process to include more flexibility for better extracting photo-$z$ information.

In Figure~\ref{fig:acc}, the purple and red lines show the accuracy curves of the hybrid and hybrid transfer networks, respectively. We notice that validation curves (purple and red dashed curves) of the two networks can reach high accuracy in early epochs, and the training stops at the 47th and 30th epochs for the hybrid and hybrid transfer networks, respectively. This indicates that the hybrid network can be optimized efficiently, and the hybrid transfer network even performs better.

\section{Result}\label{sec:result}

\begin{figure*}
    \centerline{
        \includegraphics[scale=0.26]{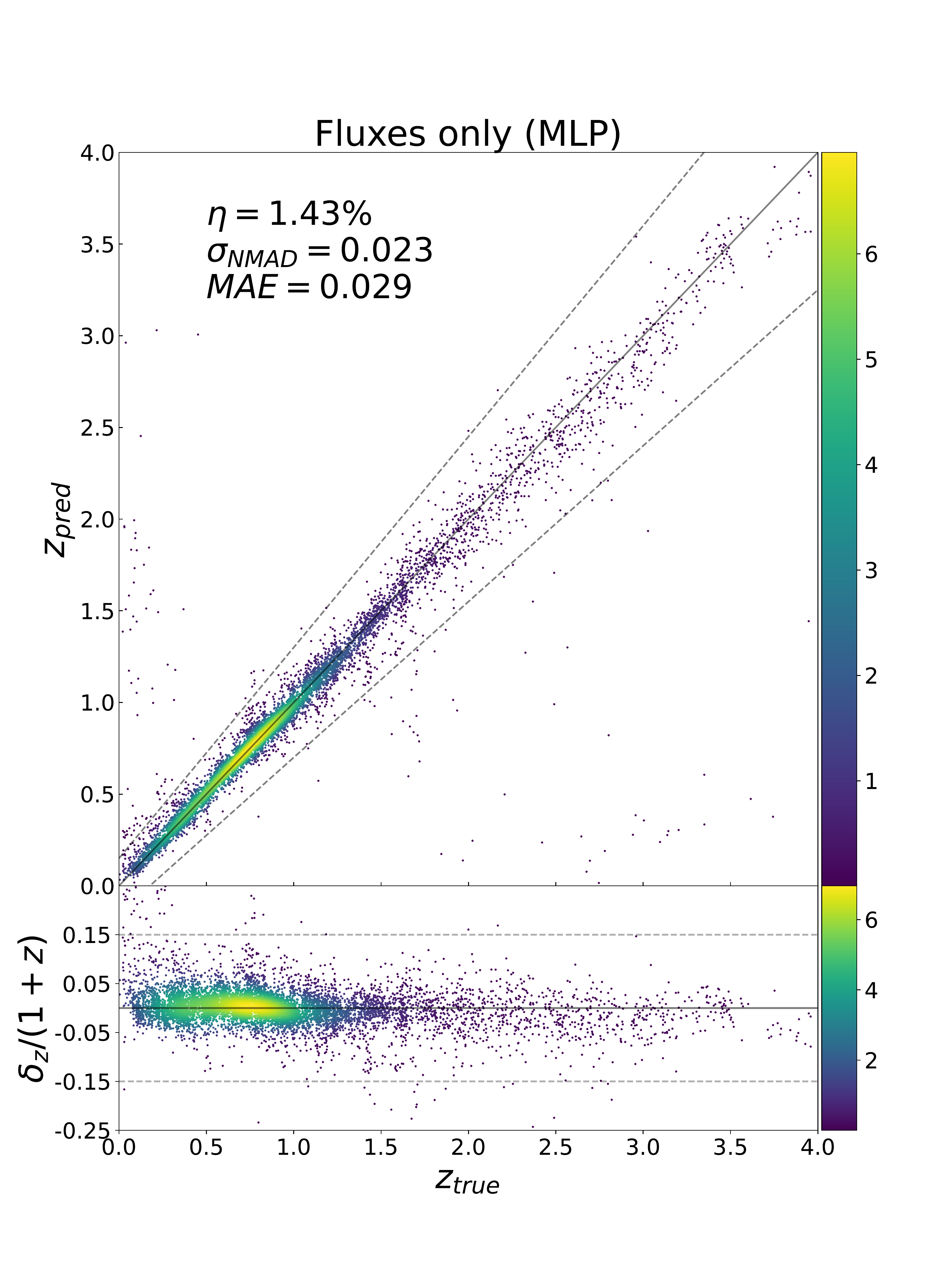}
        \includegraphics[scale=0.26]{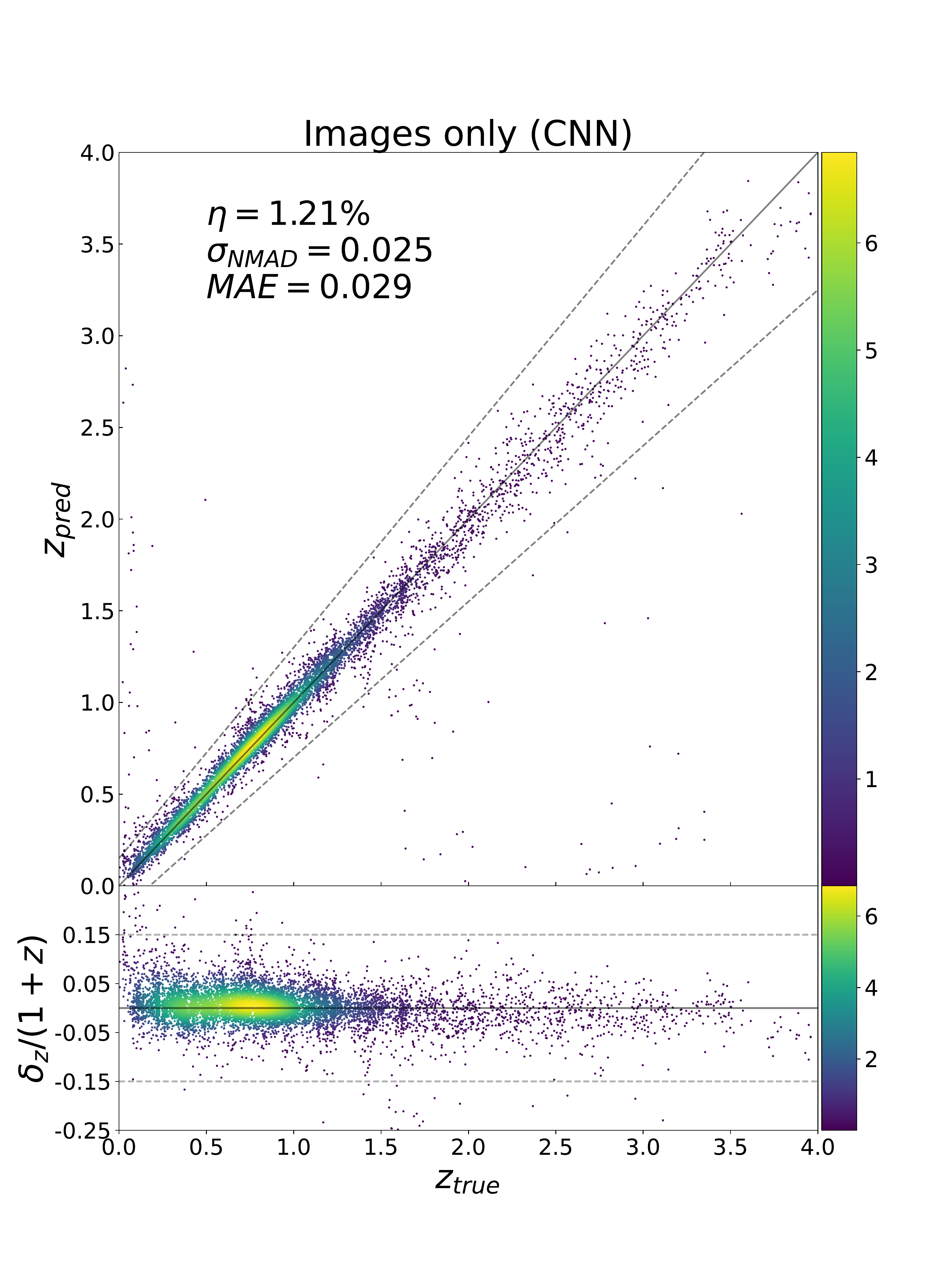}}
    \centerline{
        \includegraphics[scale=0.26]{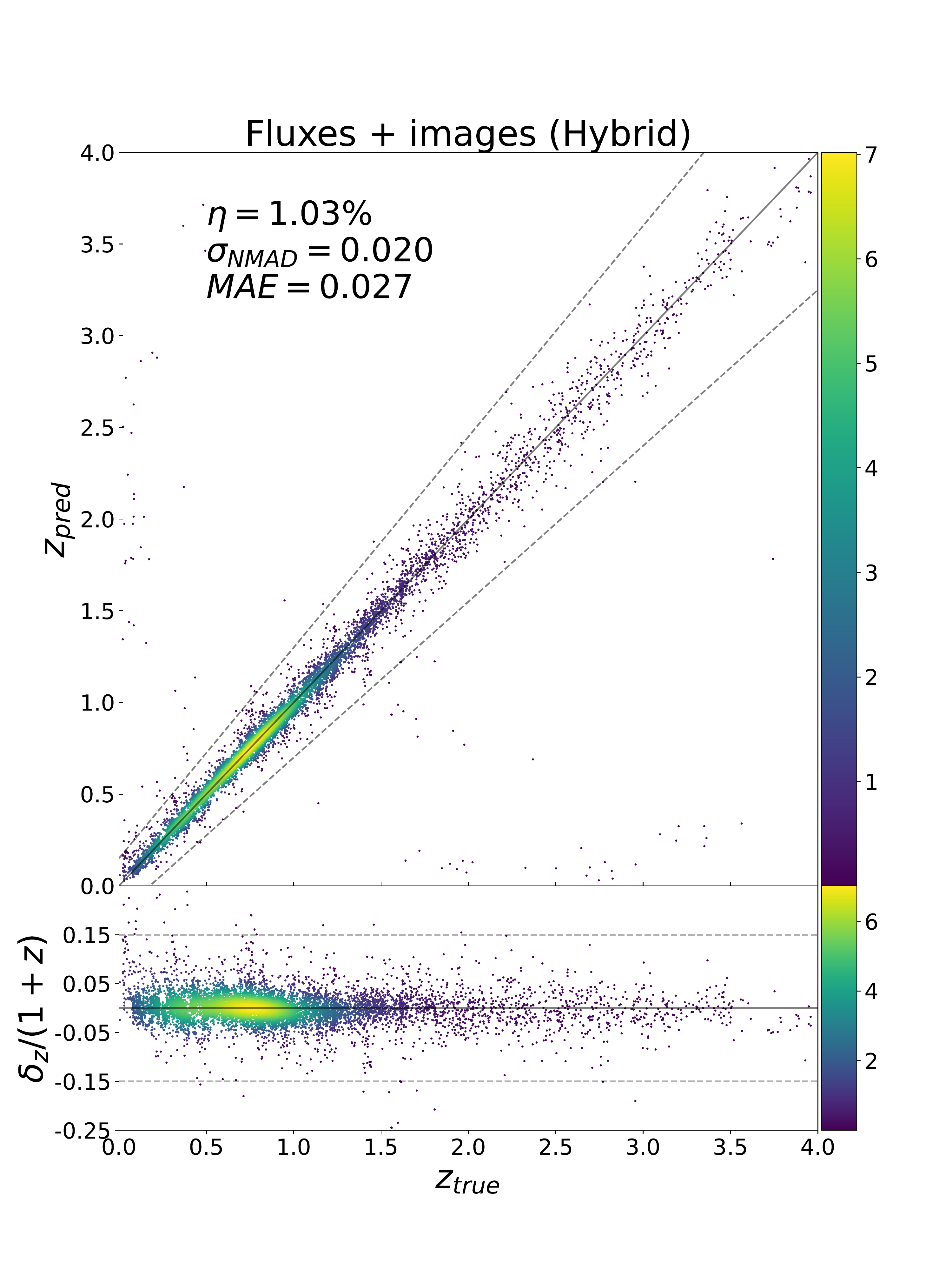}
        \includegraphics[scale=0.26]{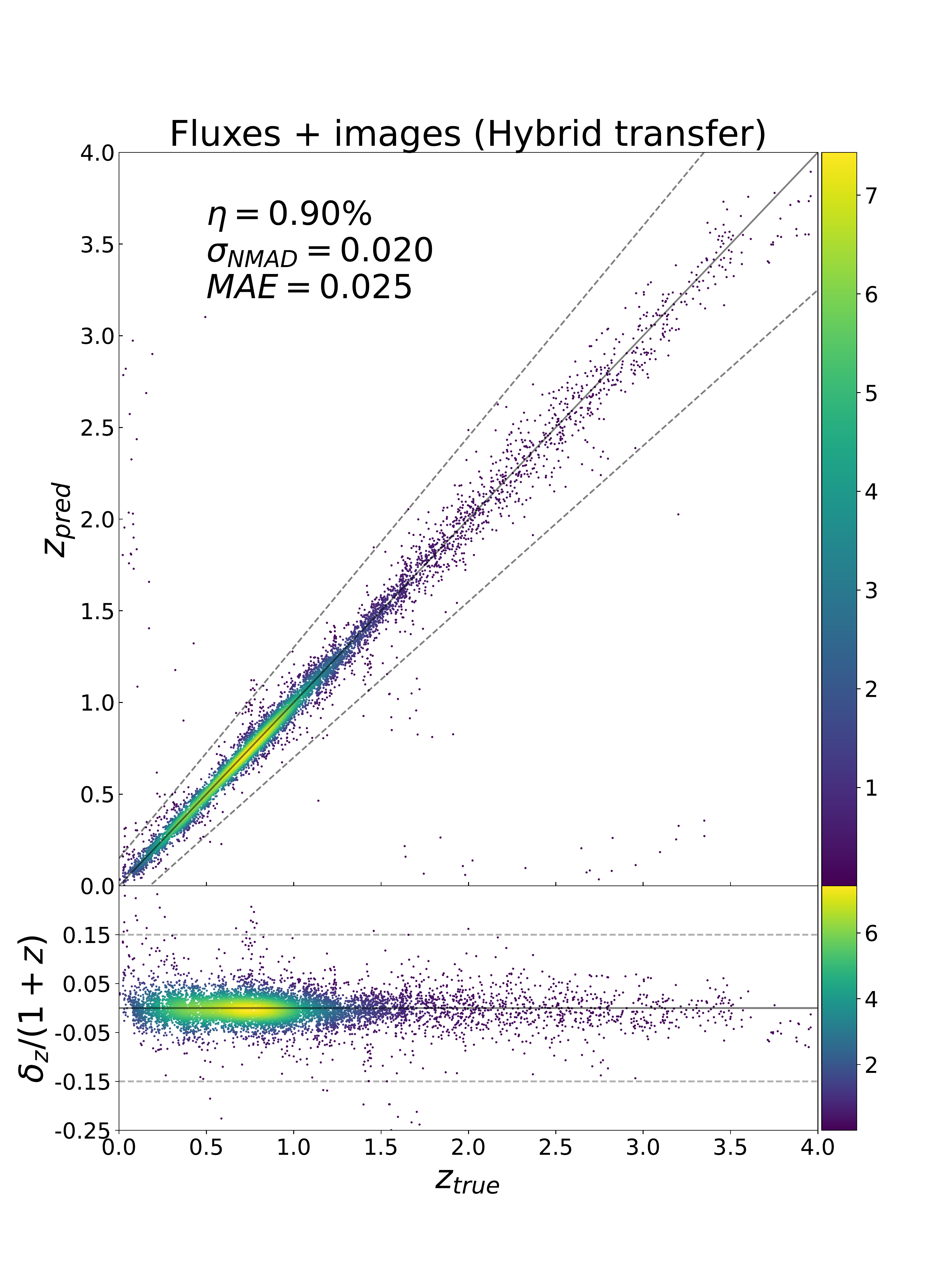}}
    \caption{\label{fig:photoz_result} Results of the MLP, CNN, hybrid and hybrid transfer networks when the ratio of training to testing data roughly equals to $3: 1$. The color bar at right of each panel denotes the number density of sources per pixel of the testing data. The MLP using flux data can provide lower photo-$z$ scatter, while the CNN using image data has smaller outlier fraction. The hybrid network can improve the results of both accuracy and outlier fraction, and the hybrid transfer network can even further suppress the outliers and has the best photo-$z$ result.}
\end{figure*}

To assess the accuracy of extracted photo-$z$ from the networks, we define catastrophic outliers to be $|\Delta z|/(1+z_{\rm true}) < 0.15$, where $\Delta z = z_{\rm pred} - z_{\rm true}$, and adopt the normalized median absolute deviation (NMAD)~\citep{Brammer2008}, which is calculated as:
\begin{equation}
    \sigma_{\rm NMAD} = 1.48 \times \rm {median}\left(\left| \frac{\Delta z - \rm{median}(\Delta z)}{1 + z_{\rm true}}\right|\right).
\end{equation}
This scatter or deviation can naturally suppress the weights of the outlier redshift, and provide a proper estimation of the photo-$z$ accuracy. We also calculate the average of $|\Delta z|/(1+z_{\rm true})$ as mean absolute error (MAE) in the analysis.

\begin{figure}
    \centering
    \includegraphics[width=\columnwidth]{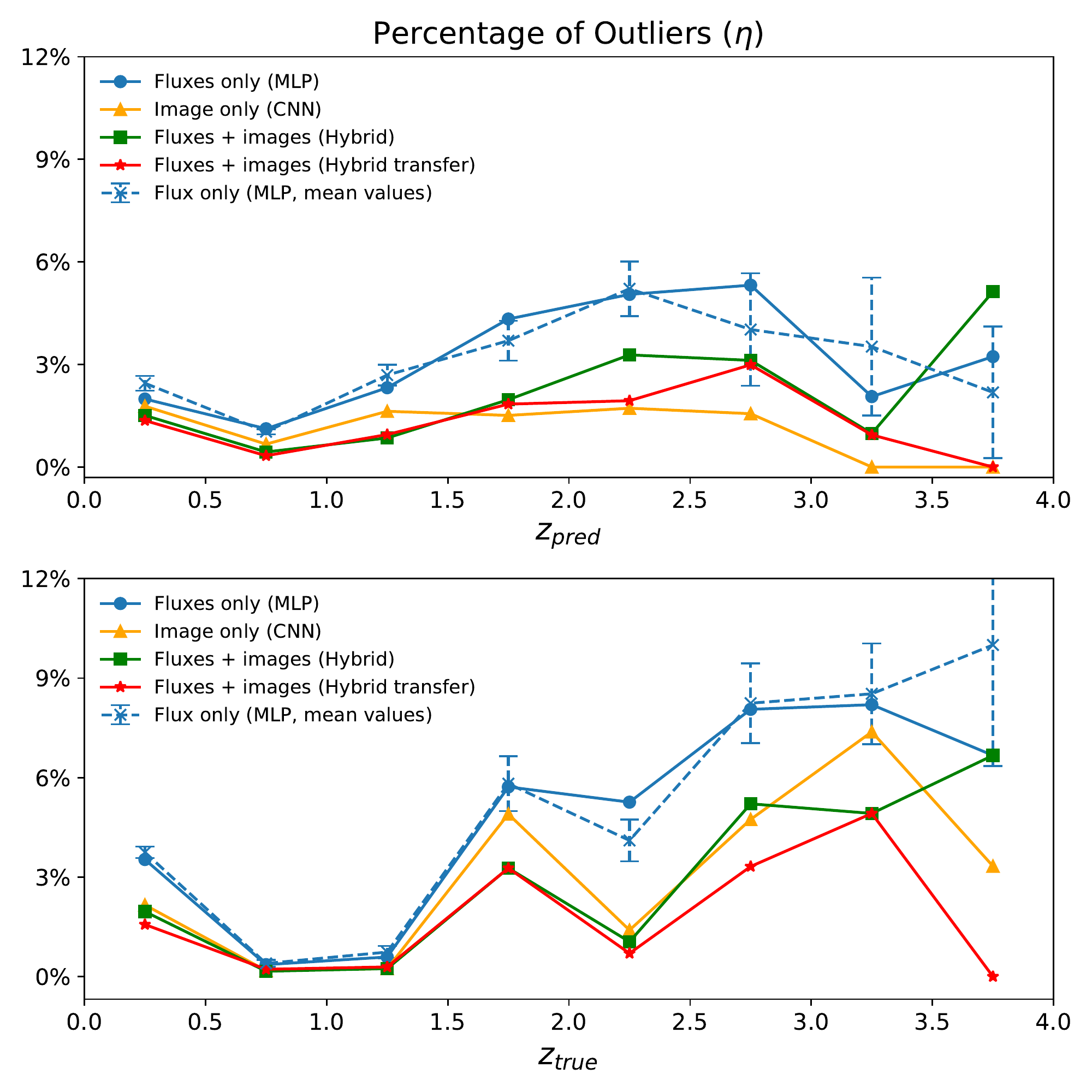}
    \caption{The percentages of outliers as a function of photometric redshift (top panel) and spectroscopic redshift (bottom panel). The blue, orange, green, and red curves are for the MLP, CNN, hybrid, and hybrid transfer networks, respectively. The average outlier fractions and errors derived from ten mock flux testing datasets using MLP are shown in blue dashed curves and data points with error bars.}
    \label{fig:outliers}
\end{figure}

In the upper-left panel of Figure~\ref{fig:photoz_result}, we show the result from the MLP using the flux data. We find that the outlier fraction $\eta = 1.43\%$, $\sigma_{\rm NMAD} = 0.023$ for the testing data. This result has been improved by a factor 2 for the outlier fraction compared to that using the SED template-fitting method shown in \cite{Cao2018}. On the other hand, the result is worse than that given in \cite{Zhou2021}. This is because in this work the flux data we use are obtained by directly measuring galaxy images with aperture photometry, instead of adopting the data estimated from SED templates used in \cite{Zhou2021}, which are more realistic and including more noises. We can see that the MLP tends to predict larger redshifts at $z\lesssim0.5$ and smaller when $z\gtrsim1.5$. We also note that predictions have little outliers in redshift range $0.5<z<1.5$, that is assuring high photo-$z$ accuracy for most of galaxies observed by the CSST (see Figure~\ref{fig:redshift distri}).

The CNN result is shown in the upper-right panel of Figure~\ref{fig:photoz_result}, and we have $\eta = 1.21\%$ and $\sigma_{\rm NMAD} = 0.025$. We can find that this result has smaller outlier fraction and a bit larger $\sigma_{\rm NMAD}$ compared to the MLP using the flux data. Since our galaxy flux data are measured from galaxy images by aperture photometry, in principle, besides the morphology information, the images also abstractly include the information of fluxes and colors. Hence, if the CNN can successfully extract all or part of these information, it is capable to obtain a comparable and even better photo-$z$ result than that using the flux data only. Based on the current result, it seems that our CNN can really extract some flux information from galaxy images in addition to morphology. In Figure~\ref{fig:photoz_result}, we can see that the outlier distribution from the CNN is similar to the MLP, but it can further suppress the outlier fraction in the whole redshift range.

In the lower-left and lower-right panels of Figure~\ref{fig:photoz_result}, we show the results from the hybrid and hybrid transfer networks, respectively. We find that our hybrid network applying transfer learning technique indeed can mildly improve the photo-$z$ estimation with $\eta= 0.90\%$ and $\sigma_{\rm NMAD} = 0.020$ compared to the hybrid network with $\eta= 1.03\%$ and $\sigma_{\rm NMAD} = 0.020$. This indicates that both our hybrid and hybrid transfer network could provide accurate photo-$z$ estimates by including galaxy flux and image data, and the hybrid transfer network performs even better with higher efficiency and accuracy. This implies that the features from trained CNN and MLP are probably better than features directly learned from Hybrid when predicting photo-$z$. In Figure~\ref{fig:photoz_result}, we notice that the hybrid transfer network can further suppress the outliers by 37\% and 26\% compared to the MLP and CNN cases, respectively, especially in $z\lesssim0.5$ and $z\gtrsim1.5$. In addition, the results from hybrid or hybrid transfer networks are obviously better than the MLP only and CNN only cases, that means the CNN indeed can extract useful morphological information from galaxy images to improve the photo-$z$ estimation.

The results shown above are obtained by using a ratio of $r=3:1$ for the training and testing data. We also test if smaller training sample can severely affect our photo-z estimation, since the network training is restricted to the spectroscopic data providing accurate redshifts and we may not have large qualified training sample in real observations. Here we try two training-testing ratios $r=1: 1$ and $1: 3$, that means 20,000 and 10,000 training data are randomly selected, respectively, and the rest are saved for testing. After training and testing processes, the networks can give $\eta=1.67\%, 1.51\%, 1.25\%, 1.06\%$ and $\sigma_{\rm NMAD} = 0.026, 0.027, 0.022, 0.021$ for the MLP, CNN, hybrid and hybrid transfer networks, respectively, for $r=1: 1$ or 20,000 training sample case. The results become $\eta=2.39\%$, 1.98\%, 1.29\% and 1.29\%, and $\sigma_{\rm NMAD}=0.028$, 0.030, 0.023, and 0.023 for $r=1: 3$ or 10,000 training sample case. We can note that decreasing training data indeed result in worse predictions, but not significantly. The hybrid and hybrid transfer network seem mostly immune to this ratio in these four networks, and still can provide $\eta\sim1\%$ and $\sigma_{\rm NMAD}\sim0.02$, that is similar as the $r=3:1$ or 30,000 training sample case. On the other hand, the results become obviously worse for the MLP and CNN. This indicates that relatively more data are needed to sufficiently train the MLP and CNN, compared to the hybrid and hybrid transfer networks. We also should note that the absolute number of training data determines the performance of trained model instead of the ratio of training and testing sample. As long as we have large enough spec-$z$ data which can represent the features of all galaxies in photometric survey, the neural networks can be well trained and applied to derive photo-$z$, no matter how large of the ratio of spectroscopic and photometric data. Besides, the size of the testing sample should also be large enough, the results otherwise can be affected by statistical effects like cosmic variance.

In Figure~\ref{fig:outliers}, the percentages of outliers as a function of photometric redshift and spectroscopic redshift are illustrated in the top and bottom panels, respectively. The redshift bin size is set to be $\Delta z=0.5$. The average outlier fractions and errors derived from ten mock flux testing datasets using MLP are also shown, which can indicate the statistical effects of number of sources in the training sample at different redshifts. Although currently it is hard for us to generate enough mock images to derive errors in the CNN case, the errors should be similar as the MLP case, since they have similar photo-$z$ estimation results. This is also representative for the cases using the hybrid and hybrid transfer networks. In the bottom panel, we can see that, generally speaking, the outlier fraction is increasing as redshift increases. In $z=0.5-1.5$, all of the four networks give the lowest outliers, that is probably due to large training sample in this range (see Figure~\ref{fig:redshift distri}). In $z=3.5-4$, there are few galaxies and the results are dominated by statistical errors, which leads to large scatters in the results between the four networks. The results can be heavily disturbed if shown as a function of photo-$z$ as indicated in the top panel. In this case, we still have small outliers in $z=0.5-1$, but it is not visibly increasing but is relatively flat as photo-$z$ increases.

\section{Summary and Conclusion}\label{sec:conclusion}

We explore the photo-$z$ accuracy of the CSST photometric observations using neural networks. Four neural networks, i.e. MLP, CNN, Hybrid and Hybrid transfer, are adopted and tested to extract the photo-$z$ information from the CSST mock flux and image data.
We first simulate the CSST photometric observations to generate mock galaxy images for the seven bands, which are created based on the HST-ACS observations and the COSMOS catalog considering the CSST instrumental effects. Then we measure fluxes and errors in each band from galaxy images using aperture photometry method. Since the photo-$z$ data would mainly be used in the analysis of the CSST weak gravitational lensing surveys, about 40,000 high-quality sources with ${\rm SNR} > 10$ in the $g$ or $i$ band are selected as the training and testing samples in our neural networks.

The MLP and CNN are used to predict photo-z from fluxes and images, respectively. Since the flux measured by aperture photometry could be negative for faint galaxies especially in the bands with low transmissions, we rescale the measured galaxy flux, error, and color, which also can speed up the training process of the networks. These flux data are inputted into the constructed MLP with six fully connected layers for deriving photo-$z$. The mock galaxy images in CSST seven photometric bands are cropped or padded into the same size as the inputs for the CNN. The inception blocks are employed to extract image features parallelly with different sizes of kernels. To investigate the improvement of photo-$z$ accuracy using both flux and image data, we develop a hybrid network by properly combining the MLP and CNN. We also propose an hybrid transfer network, adopting the transfer learning techniques, which may have higher efficiency and accuracy.

In the MLP training process, to effectively suppress statistical noises of flux data, we randomly generate 50 realizations from Gaussian distribution for each galaxy flux data based on errors as the training data. We also augment galaxy images by rotating and flipping to produce more training data by a factor of 8 in the CNN training process. In the hybrid network, we randomly input a corresponding rotating or flipping galaxy image in the CNN part given a flux realization, and can efficiently provide accurate photo-$z$ result. In the hybrid transfer network, we freeze the weights of the MLP and CNN parts as the ones separately trained by the flux and image data. We find that the MLP, CNN, hybrid and hybrid transfer networks can derive accurate photo-$z$ results with similar deviation 0.020-0.025 and outlier fraction around $1\%$. The CNN outlier fraction is better than the MLP, since our CNN can effectively extract both flux and morphology information from galaxy images. The hybrid (hybrid transfer) network can offer best result with 28\% and 15\% (37\% and 26\%) improvements compared to the MLP and CNN, respectively. The effect of training and testing ratio is also explored, and we find that smaller training sample would not affect the networks significantly.

Note that we only predict photo-$z$ values without uncertainties or probability distributions. In future work, we will try to use probabilistic deep learning methods, such as Bayesian neural networks (BNN)~\citep{MacKay1995, Neal1996, Blundell2015, Gal2015, Wilson2020} or other approaches to output both photo-$z$ values and uncertainties, and even their probability distributions.

\section*{Acknowledgements}

X.C.Z. and Y.G. acknowledge the support of MOST-2018YFE0120800, 2020SKA0110402, NSFC-11822305, NSFC-11773031, NSFC-11633004, and CAS Interdisciplinary Innovation Team. X.L.C. acknowledges the support of the National Natural Science Foundation of China through grant No. 11473044, 11973047, and the Chinese Academy of Science grants QYZDJ-SSW-SLH017, XDB 23040100, XDA15020200. L.P.F. acknowledges the support from NSFC grants 11933002, and the Dawn Program 19SG41 \& the Innovation Program 2019-01-07-00-02-E00032 of SMEC. This work is also supported by the science research grants from the China Manned Space Project with NO.CMS-CSST-2021-B01 and CMS- CSST-2021-A01.

%%%%%%%%%%%%%%%%%%%%%%%%%%%%%%%%%%%%%%%%%%%%%%%%%%
\section*{Data Availability}

The data that support the findings of this study are available from the corresponding author, upon reasonable request.



\bibliographystyle{mnras}
\bibliography{Reference} % if your bibtex file is called example.bib

\bsp	% typesetting comment
\label{lastpage}
\end{document}